\documentclass[11pt, a4paper]{article}
\usepackage{graphicx}
\usepackage[hscale=0.75,vscale=0.8]{geometry}
\usepackage{amsfonts}
\usepackage{amsmath} 
\usepackage{bm}

\begin{document}

\title{A product picture for quantum electrodynamics}

\author{Bernard S. Kay$^*$  \medskip \\  {\small \emph{Department of Mathematics, University of York, York YO10 5DD, UK}} \smallskip \\ 
\small{$^*${\tt bernard.kay@york.ac.uk}}}

\date{}

\maketitle

\begin{abstract} 
We present a short account of our work to provide quantum electrodynamics with a \emph{product picture}. It aims to complement the longer exposition in a recent paper in \emph{Foundations of Physics} and to help to make that work more accessible.  The product picture is a formulation of QED, equivalent to standard Coulomb gauge QED, but in which the Hilbert space arises as (a certain physical subspace of) a product of a Hilbert space for the electromagnetic field and a Hilbert space for charged matter (i.e.\ the Dirac field) and in which the Hamiltonian arises as the sum of an electromagnetic Hamiltonian and a charged matter Hamiltonian and an interaction term.   (The Coulomb gauge formulation of QED is not a product picture because, in it, the longitudinal part of the electromagnetic field is made out of charged matter operators.)  We also recall a `Contradictory Commutator Theorem' for QED which exposes flaws in previous attempts at temporal gauge quantization of QED and we explain how our product picture appears to offer a way to overcome those flaws.  Additionally, we discuss the extent to which that theorem may be generalized to Yang-Mills fields.  We also develop a product picture for nonrelativistic charged particles in interaction with the electromagnetic field and point out how this leads to a novel way of thinking about the theory of many nonrelativistic electrically charged particles with Coulomb interactions.   In an afterword, we explain how the provision of a product picture for QED gives hope that one will be able likewise to have a product picture for (Yang Mills and for) quantum gravity -- the latter being needed to make sense of the author's \emph{matter-gravity entanglement hypothesis}.  Also, we briefly discuss some similarities and differences between that hypothesis and its predictions and ideas of Roger Penrose related to a possible r\^ole of gravity in quantum state reduction and related to cosmological entropy.
\end{abstract}

\section{Introduction: Coulomb Gauge QED and its aesthetically unpleasant features}

Let me begin by recalling the standard Coulomb gauge formulation for a charged Dirac field, $\psi$, in interaction with the quantum electromagnetic field.   The Hamiltonian is 
\begin{equation}
\label{QEDHam}
H_{\mathrm{QED}}^{\mathrm{Coulomb}} = \int \frac{1}{2} {\bm\pi^\perp}^2\! + \frac{1}{2}({\bm\nabla} {\bm\times} {\bm A})^2 +\psi^*\gamma^0{\bm \gamma}{\bm\cdot}(-i{\bm\nabla} - {\rm e}{\bm A})\psi+m\psi^*\gamma^0\psi \, 
d^3 x + V_{\mathrm{Coulomb}}
\end{equation}
where
\[
V_{\mathrm{Coulomb}}=\frac{\rm e^2}{2} \int\int\frac{\psi^*(\bm x)\psi(\bm x)\psi^*(\bm y)\psi(\bm y)}{4\pi|\bm x-\bm y|} \, d^3x d^3y \  
\]
is the standard Coulomb action-at-a-distance electrostatic potential energy of the Dirac charge density
\begin{equation}
\label{rho}
\rho(\bm x)={\rm e}\psi^*(\bm x)\psi(\bm x).
\end{equation}
\smallskip
Here, $\rm e$ is the charge on the electron, $\bm A$ is the vector potential for the magnetic field satisfying the Coulomb gauge condition, $\bm\nabla\bm{\cdot A}=0$, and $\bm\pi^\perp$ is (minus) the transverse part (satisfying $\bm\nabla\bm{\cdot \bm\pi^\perp}=0$) of the electric field.

This is to be supplemented with the commutation and anticommutation relations
\begin{equation}
\label{CR}
[A_i(\bm x), {\pi^\perp}_j(\bm y)]=i\delta_{ij}\delta^{(3)}(\bm x-\bm y) + i\frac{\partial^2}{\partial x^i\partial x^j}
\left ({1\over 4\pi|\bm x-\bm y|}\right),
\end{equation}
\[
\lbrace\psi(\bm x), \psi^*(\bm y)\rbrace=\delta^{(3)}({\bm x}-{\bm y})
\]
while the commutators of $A_i(\bm x)$ with $A_j(\bm y)$, $\pi^\perp_i(\bm x)$ with $\pi^\perp_j(\bm y)$, and of $A_i(\bm x)$ and $\pi^\perp_i(\bm x)$ with $\psi(\bm y)$ and $\psi^*(\bm y)$ vanish, and also the anticommutator of $\psi(\bm x)$ with $\psi(\bm y)$ and of $\psi^*(\bm x)$ with $\psi^*(\bm y)$ vanish.

While it leaves unaddressed many questions of mathematical rigour (and the related difficulties of renormalization) the Hamiltonian \eqref{QEDHam} together with the commutation and anticommutation relations \eqref{CR} provide us with a suitable starting definition for what we know to be the correct theory of quantum electrodynamics (QED). For example, QED is introduced in this way in Steven Weinberg's textbook on quantum field theory \cite{Weinberg}. 

However, and leaving aside questions of mathematical rigour, this Coulomb-gauge version of quantum electrodynamics suffers from a number of aesthetically unpleasant features.   An obvious such unpleasant feature is that it makes implicit reference to a particular choice of Lorentz frame.  (The way our product picture transforms under Lorentz transformations is an interesting question but we won't address it here.)   But, leaving that aside too, we wish here to focus on the following:

\begin{enumerate}
\item $V_{\mathrm{Coulomb}}$ is quartic in $\psi$ and also nonlocal.

\smallskip

\item The commutation relations \eqref{CR} also have a nonlocal piece in addition to the delta function on the right hand side.   (One can see that such a term is necessary to be consistent with the transversality of ${\bm \pi}^\perp$ and of the Coulomb gauge condition on $\bm A$ on the left hand side.)

\smallskip

\item Only the transverse part of the electric field participates in the dynamics.   Instead of a longitudinal part, we have an action-at-a-distance force due to the potential $V_{\mathrm{Coulomb}}$.  This is unsatisfactory because, for example, it suggests that we need to think of the energy stored in a capacitor as due to the work required to separate its charged plates, or the work done against the electrostatic attraction of opposite charges to charge them up, whereas we would prefer (since Faraday!) to think it is stored in the (longitudinal) electric field between the plates!

\smallskip

\item True we can define the longitudinal electric field by $\bm E^{\mathrm{long}} = - \bm\nabla\phi$ where
\begin{equation}
\label{phi}
\phi({\bm x})= {\rm e}\int \frac{\psi^*({\bm y})\psi({\bm y})}{4\pi|{\bm x} - {\bm y}|}\,d^3{\bm y}
\end{equation}
which, by the way, is equivalent to imposing Gauss's law, $\bm{\nabla.E}=\rho$, by definition!
But then $E^{\mathrm{long}}$ is made out of Dirac field operators and doesn't have an independent existence (as well as not participating in the dynamics).

\smallskip

\item \label{LastPoint} It would be nice if the Hilbert space, ${\cal H}_{\mathrm{QED}}$, could be written as the tensor product,
\begin{equation}
\label{HilbProd}
{\cal H}_{\mathrm{QED}}= {\cal H}_{\mathrm{electromag}}\otimes {\cal H}_{\mathrm{charged \ matter}},
\end{equation}
of a Hilbert space, ${\cal H}_{\mathrm{electromag}}$, for the electromagnetic field and a Hilbert space, ${\cal H}_{\mathrm{charged \ matter}}$, for charged matter (i.e.\ for our Dirac field) and if also the QED Hamiltonian then arose in the form
\begin{equation}
\label{ProdHam}
H_{\mathrm{QED}} = H_{\mathrm{electromag}} + H_{\mathrm{charged \ matter}} + H_{\mathrm{interaction}}
\end{equation}
When such a statement is true of a quantum theory involving two systems in interaction, we say it has a \emph{product structure}.  But the Coulomb-gauge formulation of QED doesn't have a ``product structure'' in this sense.  It might appear to do so if $H_{\mathrm{electromag}}$ were interpreted as the Hilbert space of the transverse part of the electromagnetic field.   But if we insist on all electromagnetic field operators (both transverse and longitudinal) acting on ${\cal H}_{\mathrm{electromag}}\otimes {\cal H}_{\mathrm{charged \ matter}}$ as operators of the form $A\otimes I$, where $I$ denotes the identity on on ${\cal H}_{\mathrm{charged \ matter}}$  -- which is what we really mean when we say that ${\cal H}_{\mathrm{electromag}}$ is ``the Hilbert space for the electromagnetic field'' -- then this clearly doesn't hold since the longitudinal part of the electric field is made out of charged matter field operators.

\end{enumerate}

To amplify on Point \ref{LastPoint}, what would be preferable (and as we shall discuss further in Section \ref{Afterword} not only for aesthetic reasons but also because it enables one to ask and answer new physical questions) would be if there were a way to quantize electrodynamics in which the Hilbert space took the product form  \eqref{HilbProd} and in which the Hamiltonian took the 	\emph{product picture} form \eqref{ProdHam}, or more precisely, if it took the form
\begin{equation}
\label{HamPP}
H^{\mathrm{PP}}_{\mathrm{QED}} = \int {1\over 2} {\tilde{\bm\pi}}^2\! +{1\over 2}({\bm\nabla} {\bm\times} {\hat {\bm A}})^2 +\psi^*\gamma^0{\bm \gamma}{\bm\cdot}(-i{\bm\nabla} - {\rm e}{\hat{\bm A}})\psi+m\psi^*\gamma^0\psi \, d^3 x
\end{equation}
with no Coulomb potential term.  And if the electromagnetic field commutation relations took the form
\begin{equation}
\label{CRPP}
[\hat A_i(\bm x), \tilde \pi_j(\bm y)]= i\delta_{ij}\delta^{(3)}(\bm x - \bm y), \quad \lbrace\psi(\bm x), \psi^*(\bm y)\rbrace = \delta^{(3)}(\bm x-\bm y)
\end{equation}
with no nonlocal term.  And if $\tilde{\bm \pi}$ was (minus) the \emph{full} electric field operator (rather than just its transverse part).  And if Gauss's law, $\bm{\nabla.E}=\rho$, held as a genuine operator equation (rather than by definition).  And yet if this new formulation was entirely equivalent to Coulomb gauge QED.   Because after all, while we have complained that it is aesthetically unpleasant, we do know that it's physically correct! 

\medskip

I want next to show that (at a similarly mathematically non-rigorous level to that at which the Coulomb gauge theory is typically discussed) it is possible to have a formulation of QED, which we call the \emph{product picture}, equivalent to the Coulomb gauge formulation, and with all these nice features.    More precisely, the equivalence is between the Coulomb gauge formulation and our product picture theory restricted to a certain \emph{product picture physical subspace} of a full (unphysical) product picture Hilbert space which we call below the `augmented QED Hilbert space'.

\medskip

As a preliminary to that, we conclude this introduction by recalling how the operators $\bm A$ and $\bm\pi^\perp$ in Coulomb gauge QED may be represented on a Hilbert space.   In terms of the usual (say, momentum-space) annihilation and creation operators, $a_i(\bm k)$, $a_i^+(\bm k)$  on the Fock space, ${\cal F}({\cal H}_{\mathrm{one}})$ over the one particle Hilbert space, ${\cal H}_{\mathrm{one}} = L^2({\mathbb{R}}^3)^3$, we may define
\begin{equation}
\label{ApiRep}
A_i(\bm k)=\left(\frac{a^{\mathrm{trans}}_i(\bm k)+{a_i^+}^{\mathrm{trans}}(\bm k)} 
{\sqrt 2|k|^{1/2}}\right),\ {\pi^\perp}_i(\bm k)=-i|k|^{1/2}\left(\frac{{a^{\mathrm{trans}}_i
(\bm k)-{a_i^+}^{\mathrm{trans}}}(\bm k)}{\sqrt 2}\right)
\end{equation}
where $a^{\mathrm{trans}}_i(\bm k)$ is the transverse part, 
$(\delta_i^j-k_ik^j/k^2)a_j(\bm k)$ of $a_i(\bm k)$ and 
${a^+_i}^{\mathrm{trans}}(\bm k)$ the transverse part, $(\delta_i^j-k_ik^j/k^2)a^+_j(\bm k)$ of $a_i^+(\bm k)$.  

It will be important to note here that we can think about the Hilbert space on which $A_i(\bm k)$ and ${\pi^\perp}_i(\bm k)$ act in two slightly different ways.   Either we can think of it as the Fock space, ${\cal F}({\cal H}_{\mathrm{one}}^{\mathrm{trans}})$, over the subspace, ${\cal H}_{\mathrm{one}}^{\mathrm{trans}}$, consisting of transverse elements of ${\cal H}_{\mathrm{one}}$ or we can note that the full one particle Hilbert space ${\cal H}_{\mathrm{one}}$ can be written as a sum of its two orthogonal subspaces, ${\cal H}_{\mathrm{one}}^{\mathrm{trans}}$, and ${\cal H}_{\mathrm{one}}^{\mathrm{long}}$ consisting respectively of transverse and longitudinal vectors, and thus the full Fock space, ${\cal F}({\cal H}_{\mathrm{one}})$, is the tensor product, ${\cal F}({\cal H}_{\mathrm{one}}^{\mathrm{trans}})\otimes {\cal F}({\cal H}_{\mathrm{one}}^{\mathrm{long}})$, and think of the subspace on which $A_i(\bm k)$ and ${\pi^\perp}_i(\bm k)$ act as consisting of the subspace ${\cal F}({\cal H}_{\mathrm{one}}^{\mathrm{trans}})\otimes \Omega^{\mathrm{long}}$, where $\Omega^{\mathrm{long}}$ is the vacuum vector in ${\cal H}_{\mathrm{one}}^{\mathrm{long}}$.   With the latter point of view, we admit the possibility of \emph{longitudinal photons} but keep them in their vacuum state.

With the latter viewpoint, we can think of the full representation space of QED (now incorporating the Dirac field too) as what we shall call the
\emph{Coulomb gauge physical subspace (CGPS)}, ${\cal F}({\cal H}_{\mathrm{one}}^{\mathrm{trans}})\otimes \Omega^{\mathrm{long}}\otimes {\cal H}_{\mathrm{Dirac}}$, of the \emph{augmented QED Hilbert space}, which is the name we shall give to ${\cal F}({\cal H}_{\mathrm{one}})\otimes {\cal H}_{\mathrm{Dirac}}$.

\section{\label{QEDPP} A product picture for QED}

To provide QED with a product picture, the first step is to define operators, $\hat{\bm A}$ and $\tilde{\bm\pi}$, on our augmented QED Hilbert space ${\cal F}({\cal H}_{\mathrm{one}})\otimes {\cal H}_{\mathrm{Dirac}}$ by
 \[
\hat A_i(\bm k) =\left(\frac{a_i(\bm k)+{a_i^+}(\bm k)} 
{\sqrt 2|k|^{1/2}}\right), \quad \tilde{\bm\pi} = \bm\pi^\perp + \tilde{\bm\pi}^{\mathrm{long}}
\] 
where $\bm\pi^\perp$ is defined as in Equation \eqref{ApiRep}
and
\[
\tilde\pi^{\mathrm{long}}_i(\bm k) = -\frac{2i|k|^{1/2}}{\sqrt{2}}a_i^{\mathrm{long}}(\bm k).
\]
This last definition may seem strange.  Let us pause to notice that we could think of $\tilde{\bm\pi}^{\mathrm{long}}$ as the result of first defining an operator that we shall call $\hat{\bm\pi}$ by again taking its transverse part to be the same as the definition of $\bm\pi^\perp$ in Equation \eqref{ApiRep} and defining its longitudinal part, along similar lines, i.e.\ by
\[
\hat\pi^{\mathrm{long}}_i(\bm k)=-i|k|^{1/2}\left(\frac{{a^{\mathrm{long}}_i
(\bm k)-{a_i^+}^{\mathrm{long}}}(\bm k)}{\sqrt 2}\right),
\]
but then deleting the second term on the right hand side containing a creation operator and doubling the term on the right hand side containing an annihilation operator.   Clearly, in consequence (and unlike $\hat{\bm A}$ and $\hat{\bm\pi}$) $\tilde{\bm\pi}$ will fail to be (even formally) self-adjoint.   However it will still have the same commutation relations with all other operators (and in particular with $\hat{\bm A}$) that $\hat{\bm\pi}$ has and we shall soon see (see Theorem 1a below) that it \emph{will} be (formally) self-adjoint on the appropriate  `product picture physical subspace' of the augmented QED Hilbert space which we next define.

First we define the (formally) unitary operator
\begin{equation}
\label{U}
U=\exp\left(i\int \hat A^i(\bm x)\partial_i\phi(\bm x)\, d^3x\right ) 
\end{equation}
where $\phi$ is given by Equation \eqref{phi} and then we define the \emph{product picture physical subspace (PPPS)} to be the result of acting with $U$ on the Coulomb gauge physical subspace, i.e.\ as $U{\cal F}({\cal H}_{\mathrm{one}}^{\mathrm{trans}})\otimes \Omega^{\mathrm{long}}\otimes {\cal H}_{\mathrm{Dirac}}$.  Let us notice here that all states in the PPPS (including the vacuum state, $U\Omega$) are entangled between charged matter and longitudinal photons!

We then have:

\medskip

\noindent
{\bf Theorem 1a.} \emph{$\tilde{\bm\pi}$ and $H^{\mathrm{PP}}_{\mathrm{QED}}$ (defined as in \eqref{HamPP} with the above definitions for $\hat{\bm A}$ and $\tilde\pi$) map the PPPS to itself and are self-adjoint when restricted to it.}

\medskip

\noindent
{\bf Theorem 1b.} \emph{Also the magnetic field operator, $\bm B ={\bm \nabla\times}\hat{\bm A}$, maps the PPPS to itself.  (Although the [as always, unphysical] vector potential, $\hat{\bm A}$ does not.)}

\bigskip

\noindent
{\bf Theorem 2.} \emph{$H^{\mathrm{PP}}_{\mathrm{QED}}$ on the PPPS is (unitarily) equivalent to $H_{\mathrm{QED}}^{\mathrm{Coulomb}}$ on the CGPS.}

\bigskip

\noindent
{\bf Theorem 3.}  \emph{$\forall\, \bm\Psi\in$ PPPS, $\bm\nabla\bm\cdot\bm E\,\bm\Psi \ (=-\bm\nabla\bm\cdot\tilde{\bm\pi}\,\bm\Psi) =\rho\bm\Psi$,
i.e.\ Gauss's law holds as an operator equation on the PPPS (in contrast to Coulomb gauge, where it held by definition).}

\medskip

We briefly sketch the proofs.  For more details, see \cite[Section 3]{KayQES}. 

\noindent
\emph{Proofs of Theorems 3 and 1a.} These follow easily after noting that $U\tilde{\bm\pi} U^{-1}=\tilde{\bm\pi}-\bm\nabla\phi$

\medskip

\noindent
\emph{Proof of Theorem 1b.} This follows immediately from the fact that $\bm{\nabla\times}\hat{\bm A}$ is the same as  $\bm{\nabla\times}$ acting on the transverse part of $\hat{\bm A}$ (which is the same thing as the Coulomb gauge $\bm A$) and that, as one easily sees, this maps the PPPS to itself.   (It is noteworthy,  that the longitudinal part of $\hat{\bm A}$ \emph{doesn't} map the PPPS to itself but this is annihilated by $\bm{\nabla\times}$.)

\medskip

\noindent
\emph{Proof of Theorem 2.} First define 
\begin{equation}
\label{Hcheck}
\check H_{\mathrm{QED}} = H_{\mathrm{QED}}^{\mathrm{Coulomb}} + \int \frac{1}{2}(\tilde{\bm\pi}^{\mathrm{long}})^2 + \tilde{\bm\pi}\cdot\bm\nabla\phi \, d^3x
\end{equation}
and notice that, restricted to the CGPS, $\check H_{\mathrm{QED}}=H_{\mathrm{QED}}^{\mathrm{Coulomb}}$.

\bigskip

Then on the full augmented QED Hilbert space, if one calculates $U\check H_{\mathrm{QED}}U^{-1}$ one finds that it equals $H_{\mathrm{QED}}^{\mathrm{PP}}$.

\medskip

To do this calculation, it is helpful as a first step to notice that
\[
U\tilde{\bm\pi} U^{-1}=\tilde{\bm\pi}-\bm\nabla\phi, \quad U\psi U^{-1}=\breve\psi 
\]
where we define 
\[
\breve\psi = e^{ - \left (ie \frac{\partial_i}{\nabla^2}\hat A^i\right )}\psi. 
\]
Using these, one easily finds that $U\check H_{\mathrm{QED}}U^{-1}$ is equal to
\[
\int \frac{1}{2}\tilde{\bm\pi}^2 +{1\over 2}({\bm\nabla} {\bm\times} {\bm A})^2 + \breve\psi^*\gamma^0{\bm \gamma}{\bm\cdot}(-i{\bm\nabla} - {\rm e}{\bm A})\breve\psi+m\breve\psi^*\gamma^0\breve\psi  \, d^3x.
\]
It is then not difficult to see (or see \cite[Section 3]{KayQES}) that this is equal to the $H_{\mathrm{QED}}^{\mathrm{PP}}$ of \eqref{HamPP}.  (End of proofs.)   

Let us also remark that one may think of $\breve\psi$ as the appropriate counterpart to $\psi$ in the product picture.  In fact we have the dictionary shown in the table.

\smallskip

\begin{table}[ht]
\label{table}
\caption{\textbf{Coulomb gauge to product picture dictionary}}

\smallskip

\centering 

\resizebox{\textwidth}{!}{\begin{tabular}{c c c}
\hline 
Quantity & Coulomb gauge & product picture 
\\ [0.5ex] 
\hline
\hline
\quad & \quad & \quad \\
Hamiltonian & $\check H_{\mathrm{QED}}$   \ (or $H_{\mathrm{QED}}^{\mathrm{Coulomb}}$)  & $H^{\mathrm{PP, Dirac}}_{\mathrm{QED}}$\\
electric field & $\bm E_{\mathrm C} =  -(\tilde{\bm\pi}+\bm\nabla\phi)$ \ (or $ -(\bm\pi^\perp + \bm\nabla\phi)$)  &  ${\bm E}_{\mathrm{PP}} = -\tilde{\bm\pi}$\\ 
magnetic field & $\bm\nabla\bm\times\bm A$ \ (or $\bm\nabla\bm\times{\bm\hat A}$) & $\bm\nabla\bm\times\hat{\bm A}$ \\
Dirac field & $\psi$ &  $\breve\psi = e^{ - \left (ie \frac{\partial_i}{\nabla^2}\hat A^i\right )}\psi$ \\
adjoint Dirac field & $\psi^*$ & $\breve\psi^*= e^{\left (ie \frac{\partial_i}{\nabla^2}\hat A^i\right )}\psi^*$ \\
Dirac electrical potential & $\phi$ & $\phi$\\
vacuum state & $\Omega^{\mathrm{trans}}\otimes\Omega^{\mathrm{long}}\otimes \Omega_{\mathrm{Dirac}}$ \ (or $\Omega^{\mathrm{trans}}\otimes \Omega_{\mathrm{Dirac}}$) \quad\quad & \quad\quad $\Omega^{\mathrm{trans}}\otimes U(\Omega^{\mathrm{long}}\otimes \Omega_{\mathrm{Dirac}})$ (entangled)\\
\quad & \quad & \quad \\
\hline
\end{tabular}}
\label{tab2} 
\end{table}

Finally let us remark that there are two ways that we may obtain the unitary time evolution $\exp(-iH_{\mathrm{QED}}^{\mathrm{PP}} t)$ on the PPPS (which must both lead to the same answer).   One way is to exponentiate ($-i$ times) the self-adjoint restriction of $H_{\mathrm{QED}}^{\mathrm{PP}}t$ to the PPPS on the PPPS.    But another way would be to first exponentiate ($-i$ times) the non-self-adjoint $H_{\mathrm{QED}}^{\mathrm{PP}}t$ on the augmented QED Hilbert space -- obtaining a nonunitary operator -- and then to restrict that nonunitary operator to the PPPS on which it will of course be unitary.  In an approach to constructing QED in a mathematically fully meaningful way based on the product picture, the latter method may be technically advantageous, in part because it would enable us to separate the issue of defining the dynamics from the issue of assigning a mathematical meaning to the operator $U$ of Equation \eqref{U} in Section 2 which defines the PPPS.   This seems worthy of further exploration.

\section{The Contradictory Commutator Theorem and comparison with temporal gauge quantization}

The Hamiltonian, \eqref{HamPP}, resembles the classical Hamiltonian appropriate to the \emph{temporal gauge} condition $A^0=0$.  (Also known as the Weyl gauge condition.)  Several attempts have been made in the past to obtain a formulation of QED by `quantizing' that temporal gauge classical Hamiltonian -- taking the commutation relations to be those of Equation \eqref{CRPP} but hitherto certain assumptions have been (sometimes tacitly) made that led to pathologies of one sort or another.  (See, for example, \cite{LMS}.)
For example one could define $\hat{\bm A}$ in \eqref{CRPP} as we have here but replace $\tilde{\bm\pi}$ by the $\hat{\bm\pi}$ that we mentioned in Section \ref{QEDPP}.  But one finds that any attempt to obtain a theory equivalent to Coulomb gauge QED with those choices would fail.  The essence of what goes wrong can be summed up in what we shall call the Contradictory Commutator Theorem. (See \cite[Section 3.4]{KayQES} for more details.)

\medskip

\noindent
{\bf The Contradictory Commutator (CC) Theorem.} \emph{There can be no pair of 3-vector operators $\bm{\mathsf A}$ and $\bm{\mathsf\pi}$ on a Hilbert space $\mathsf H$ such that} 

\smallskip

\noindent
(a) \emph{$\bm{\mathsf A}$ and $\bm{\mathsf\pi}$ satisfy the canonical commutation relations}
\begin{equation}
\label{CCR}
[{\mathsf A}_i(\bm x), {\mathsf\pi}_j(\bm y)]=i\delta_{ij}\delta^{(3)}(\bm x-\bm y), \ [{\mathsf A}_i(\bm x), {\mathsf A}_j(\bm y)] =0= [{\mathsf\pi}_i(\bm x), {\mathsf\pi}_j(\bm y)]
\end{equation}

\noindent
(b) \emph{$\bm{\mathsf A}$ and $\bm{\mathsf\pi}$ are each self-adjoint;} 

\noindent
(c)  \emph{For some vector $\bm\Psi\in \mathsf H$ ($\bm\Psi \ne 0$)}
\[
\bm{\nabla\cdot{\mathsf\pi}}\bm\Psi= - \rho\bm\Psi
\]
\emph{for some operator-valued function of $\bm x$, $\rho$.}

\noindent
(d)  \emph{$\rho$ commutes with $\bm{\mathsf A}$.}

\medskip

\noindent
\emph{Proof.} (a) easily implies the equality
\begin{equation}
\label{contraeq}
\langle\bm\Psi| [{\mathsf A}_i(\bm x), \bm{\nabla\cdot{\mathsf\pi}}(\bm y)] \bm\Psi\rangle
=-i(\nabla_i\delta^{(3)})(\bm x-\bm y),
\end{equation}
while (b), (c) and (d) imply that quantity on the left hand side of \eqref{contraeq} is zero -- a contradiction!  

\medskip

However our product picture evades the conclusions of this seeming no-go theorem, as of course it must in view of Theorem 2.   First let us notice that, in attempting to apply the CC Theorem to our product picture, there are two different interpretations we could make for the Hilbert space $\mathsf H$. We could identify it with the full augmented QED Hilbert space, but then the CC Theorem fails to apply because $\tilde{\bm\pi}$ fails to be self-adjoint on $\mathsf H$.   Alternatively, we could identify $\mathsf H$ with the product picture physical subspace,  and then $\tilde{\bm\pi}$ will map $\mathsf H$ to itself and be self adjoint on it.  However $\hat{\bm A}$ will fail to map $\mathsf H$ to itself!  So either way the Contradictory Commutator Theorem fails (as it must) to rule out our product picture formulation of QED.

\subsection{The question of whether and to what extent the Contradictory Commutator Theorem generalizes to Yang-Mills}

It is natural to ask to what extent the formulation and proof of  the Contradictory Commutator Theorem generalize to Yang-Mills theory (say, for a compact connected semisimple Lie group with Lie algebra generators, $T^l$, satisfying $[T^l, T^m] = if^{lmn}T^n$ for totally antisymmetric structure constants $f^{lmn}$ and ${\rm tr}(T^mT^n) = \tfrac{1}{2}\delta^{mn}$)  and also to gravity in general relativity.  

We content ourselves here with a discussion of Yang-Mills theory (in flat spacetime).  Adopting signature $(+, -,-,-)$, we may take the classical Lagrangian density of pure Yang-Mills to be $-\tfrac{1}{2}{\rm tr}(F^{ab}F_{ab})$ where $F_{ab}$ is $\nabla_a A_b - \nabla_b A_a -ig[A_a, A_b]$ where $g$ is the Yang-Mills coupling constant and $A_a = A_a^lT^l$.    Choosing temporal gauge where $A_0=0$, the conjugate momentum, say $\pi_i$, to $A_i$ is $-F_{0i} = -\dot A_i$ while the counterpart to the Gauss law constraint is 
\begin{equation}
\label{YMGauss}
D_i\pi_i = -\rho
\end{equation}
where the covariant derivative, $D_i\pi_i$, written out in full, is given by
\begin{equation}
\label{Di}
\nabla_i\pi_i - ig [A_i,\pi_i].
\end{equation}
For, say, a quark-like field (assumed of course to commute with the Yang-Mills fields) with values, $|\psi \rangle$, in the (finite dimensional) Hilbert space of (say) the fundamental representation of the Lie algebra, $\rho$ is the operator on this Hilbert space given by the formula (cf.\ \eqref{rho}) 
\[
\rho = g|\psi^*\rangle\langle\psi|.
\]
The hypotheses of the CC theorem stated above for QED clearly have counterparts for Yang-Mills fields which may be expressed with the same equations provided one makes two changes in their interpretation.
First we need to reinterpret $\bm{\mathsf A}$ and $\bm{\mathsf\pi}$ as Lie-algebra valued quantum (vector) fields where it is to be understood that (now identifying our Lie algebra with its adjoint representation) next to the `$\delta_{ij}$', there is an unwritten identity operator (on the Lie algebra thought of as the representation vector space of the adjoint representation -- see below for a more accurate statement) on the right hand side of the commutation relations in Part (a).

Secondly, the equation $\bm{\nabla\cdot{\mathsf\pi}}\bm\Psi= - \rho\bm\Psi$ of Part (c) needs, in light of \eqref{YMGauss}, to be replaced by $D_i\mathsf\pi_i\bm\Psi= - \rho\bm\Psi$ for a Lie-algebra valued $\rho$ which commutes with $\bm{\mathsf A}$.  (And the equation is still assumed to hold for \emph{some} $\bm\Psi$ in the Hilbert space.)

With this changed interpretation of its hypotheses to make them relevant to Yang-Mills fields, the question we want to ask, and will partly answer, is to what extent the statement and proof of the theorem do -- or don't -- go through as before.   We will also comment on the relation between our work and the work in \cite{Dimock96} and \cite{LMS}. 

One big difference in the nonabelian case is that, because of the nonlinearity of Yang-Mills theory, one now has to deal with products of fields at the same point; notably in the second term in \eqref{Di}.  However, adopting the point of view of the first few pages of a paper \cite{Dimock96} of Dimock, one may proceed by formally manipulating expressions involving suitably smeared versions of our fields.  To begin to explain this, let us momentarily backtrack and notice that we might have re-expressed the commutation relations \eqref{CCR} in the electromagnetic case in terms of the smeared fields, $\bm{\mathsf A}(\bm u)$ and $\bm{\mathsf\pi}(\bm v)$, where $\bm u$ and $\bm v$ are 3-vector valued say smooth compactly supported test functions, which are to be interpreted in terms of the formal equations
\[
\bm{\mathsf A}(\bm u) = \int {\mathsf A}_i(\bm x)u_i(\bm x)\,d^3{\bm x}, \quad \bm{\mathsf \pi}(\bm v) = \int {\mathsf \pi}_i(\bm x)v_i(\bm x)\,d^3{\bm x}
\] 
whereupon the first commutator in Equation \eqref{CCR} would have taken the form
\begin{equation}
\label{SmearCCR}
[\bm{\mathsf A}(\bm u), \bm{\nabla\cdot\mathsf\pi}(v)] 
=-i(\bm u|\bm{\nabla\cdot v})
\end{equation}
where the symbol $(\,\cdot\,|\,\cdot\,)$ denotes the $L^2$ inner product in the space of 3-vector fields.  

Finally, let us remark that, since the unsmeared equations \eqref{contraeq} needed a `$\delta_{ij}$' on the right hand side due to the quantum vector fields being expressed as triples of quantum scalar fields in some coordinate system, the Hilbert space involved in that unsmeared version should have strictly been taken to be the result of taking the tensor product of the Hilbert space relevant to the smeared version with the (complexified) Hilbert space of triples of numbers with the inner product the (complexified) dot product.

Returning to the Yang-Mills case, we proceed likewise but (following \cite{Dimock96}) now smearing $\mathsf A_i$ with a test triple $u_i$ whose components take their values in the Lie algebra to obtain a smeared field $\bm{\mathsf A}(\bf u)$ etc.\ where we think of the latter as ``$({\mathsf A}_i|u_i)$'' (summed over $i$ from 1 to 3) where the inner product, $(\,\cdot\,|\,\cdot\,)$ now means the integral over $\mathbb{R}^3$ of minus the Killing form of ${\mathsf A_i}$ and $u_i$ (summed over $i$).  (And similarly for $\bm{\mathsf \pi}(\bf v)$.)  And let us note, related to our above remark about the `$\delta_{ij}$', that when we reinterpreted the hypotheses of our CC Theorem, written in their unsmeared version, to apply to Yang-Mills theories, then, strictly, we should have thought of the Hilbert space as the tensor product of the Hilbert space involved in the smeared version with the (again finite dimensional) Hilbert space consisting of the complexified Lie Algebra (thought of as a complex finite dimensional vector space) equipped with that inner product.  And the identity operator that we mentioned above should really have been regarded as the identity operator on that latter Hilbert space.  (This is the more accurate statement about that identity operator that we promised to make above.)

Assuming that the standard Lie-Algebra identity $([X,Y]|Z) = (X|[Y,Z])$ applies also when some of $X$, $Y$, $Z$ are quantum operators, one easily sees formally (again as in \cite{Dimock96} but adapted to our conventions) that the smeared quantity $(D_i{\mathsf\pi}_i|\bm u)$ can be equated with  $(\bm{\mathsf \pi}|(\bm{\nabla\cdot u}) - g(\bm{\mathsf A}|[\bm{\mathsf \pi}, \bm u])$ (or alternatively with $(\bm{\mathsf \pi}|(\bm{\nabla\cdot u}) + g(\bm{\mathsf \pi}|[\bm{\mathsf A},\bm u])$).   Making use of this, one easily computes formally (now writing $\bm{\mathsf A}(\bm u)$ in place of $(\bm{\mathsf A}| \bm u)$ etc.) that (cf.\ \cite[Equation (8)]{Dimock96})
\begin{equation}
\label{Dimock8}
[\bm{\mathsf A}(\bm u), D_i\mathsf\pi_i(\bm v)] =  -i(\bm u|\bm{\nabla\cdot v}) - ig\bm{\mathsf A}([\bm u, \bm v]).
\end{equation}

This is clearly the Yang-Mills counterpart to the commutator \eqref{SmearCCR}.  But a crucial difference is that, unlike 
the right hand side of \eqref{SmearCCR}, the right hand side of \eqref{Dimock8} is no longer just a c-number but there is also an additional term that involves the operator $\bm{\mathsf A}$.   So, when we take the expectation value of the above smeared commutator in a state vector, $\bm\Psi$, in the appropriate Hilbert space, $\mathsf H$, the result will depend on $\bm\Psi$ and we will no longer obtain an immediate contradiction as we did in the proof of the CC Theorem in the electromagnetic case.    Instead, let us first shift our viewpoint slightly and notice that one way of restating the CC Theorem in that electromagnetic case is that, if (following \cite{Dimock96}) we define a \emph{physical state} to be a state vector that satisfies $\bm{\nabla\cdot\pi\,\Psi}=0$, then the set of physical states is empty!   If we now attempt to adapt the steps of that proof to the Yang-Mills case, by calculating the expectation value of the commutator in the left hand side of  \eqref{Dimock8} in the two different ways (one way using Property (a), the other using Properties (b), (c) and (d)) we will clearly find, now defining a physical state vector to be one satisfying $D_i\pi_i\,\bm\Psi = 0$ that if a state vector, $\bm\Psi$, is physical then we must have that
\begin{equation}
\label{PsiAPsi}
\langle\bm\Psi|\bm{\mathsf A}([\bm u, \bm v])\bm\Psi\rangle= -g^{-1}(\bm u|\bm{\nabla\cdot v}). 
\end{equation}
Before we consider the implications of this, let us first notice that, in the Yang-Mills case, there is also a non-trivial commutator between $\bm{\mathsf \pi}(\bm u)$ and $D_i\mathsf\pi_i(\bm v)$ (cf.\ \cite[Equation (9)]{Dimock96})
\begin{equation}
\label{Dimock9}
[\bm{\mathsf \pi}(\bm u), D_i\mathsf\pi_i(\bm v)] =   i\bm{\mathsf \pi}([\bm u, \bm v]).
\end{equation}
from which, by a similar argument, we may conclude that, for a state vector, $\bm\Psi$, to be physical we must
also have that, for all Lie-Algebra valued smearing functions, $\bm w$,
\begin{equation}
\label{PsipiPsi}
\langle\bm\Psi|\bm{\mathsf \pi}(\bm w)\bm\Psi\rangle =0.
\end{equation}
(Here, we have used the fact that, in a semisimple Lie Algebra, every element arises as the Lie bracket of two other elements.)

So we have the striking result that, for a state vector to be physical, the expectation value of the field momentum $\bm{\mathsf \pi}$, which (in components) is our quantized $-F_{0i}$ ($= -\dot A_i)$, must vanish!   On the other hand, returning to the implications of \eqref{PsiAPsi}, we clearly at least have the immediate corollary that, for a state vector to be physical, the expectation value of $\bm{\mathsf A}$ must \emph{not} vanish!  More information would seem to be extractable from \eqref{PsiAPsi} but we will not attempt to do so here.

The above results certainly seem to suggest that, in the case of Yang-Mills, the set of physical state vectors (defined to be state vectors on which the Yang-Mills counterpart to Gauss's law in temporal gauge holds) is strangely restricted.   But this seems as far as one can easily go in directly generalizing our CC theorem proof for nonabelian Yang-Mills.   Note though that, in the case of linearized (nonabelian) Yang-Mills, the right hand side of \eqref{Dimock8} clearly becomes a c-number and so the proof goes through exactly as in the electromagnetic case.  Thus for linearized Yang-Mills we may conclude that the set of physical states is empty!

Returning to full (nonlinear) Yang-Mills, in the later pages of Dimock's paper, defining $J(h)$ to be the Gauss law operator $D_i\pi_i$ smeared with a test function $h$, he finds that $\bm{\mathsf{A}}$, $\bm{\mathsf{\pi}}$ and $J$ form an infinite dimensional Lie Algebra.  Formally exponentiating that to obtain an infinite dimensional Lie group, he points out that the elements which formally correspond to $e^{iJ(h)}$ may be understood as the generators of gauge transformations.   The notion of `physical state' is then redefined to mean a state annihilated by all those elements.   He then studies Hilbert space representations of this Lie Group and finds classes of representations for which it turns out that the set of physical state vectors, defined in that way, is empty!  However, using C* algebra techniques, he then finds a representation which is a certain limit of one of those classes of representations in which it is not empty.   Presumably, in the electromagnetic case, this coincides with the representation of the formal exponentiation of Equations \eqref{CCR} that was found in \cite{LMS} in which Gauss's law holds strongly but the fields $\bm{\mathsf{A}}$ and $\bm{\mathsf{\pi}}$ do not exist.  (Rather only their formal exponentials exist.)

On the other hand, for the electromagnetic case, we found here and in \cite{KayQES} that, at least at our level of mathematical rigour, our product picture has the advantage of evading the conclusions of the CC Theorem and in it the fields $\bm{\mathsf{A}}$ and $\bm{\mathsf{\pi}}$ exist.   Thus it would seem to be of interest to find a suitable counterpart to our electromagnetic product picture both for (linearized and full) Yang-Mills theory and also (for reasons discussed further in Section 5) at least for linearized quantum gravity and, as far as is possible, for full quantum gravity.   And it would seem to be of interest to attempt to make these product pictures mathematically rigorous.

\section{\label{NonRelProd} A product picture for Maxwell-Schr\"odinger theory and the new alternative understanding it entails for the nonrelativistic quantum mechanics of many charged particles}

There's a similar product picture in the quantum electrodynamics of many non-relativistic Schr\"odinger particles interacting with the EM field.   (For more details see \cite[Section 4]{KayQES}.)  One has to treat the particles as finite radius balls with charge density $\rho$.   The Hamiltonian
\[
H^{\mathrm{Coulomb\, Schr}}_{\mathrm{QED}} = \int {1\over 2} {\bm\pi^\perp}^2 +{1\over 2}({\bm\nabla} {\bm\times} {\bm A})^2 d^3x  +  \sum_{I=1}^N \frac{\left(\bm p_I - \int {\bm A}(\bm x)\rho_I(\bm x - \bm x_I) \, d^3\bm x\right)^2}{2M_I} +  V_{\mathrm{Coulomb}}^{\mathrm{Schr}}
\]
\[
\hbox{where}\quad V_{\mathrm{Coulomb}}^{\mathrm{Schr}}=\frac{1}{2}\sum_{I=1}^N\sum_{J=1}^N \int\int\frac{\rho_I(\bm x-\bm x_I)\rho_J(\bm y-\bm x_J)}{4\pi|\bm x-\bm y|}\, d^3x d^3y
\]
on the appropriate CGPS gets replaced by
\begin{equation}
\label{HamPPSchr}
H^{\mathrm{PP\, Schr}}_{\mathrm{QED}} = \int {1\over 2} \tilde{\bm\pi}^2 +{1\over 2}({\bm\nabla} {\bm\times} \hat{\bm A})^2 \, d^3x + \sum_{I=1}^N \frac{\left(\bm p_I - \int \hat{\bm A}(\bm x)\rho_I(\bm x - \bm x_I) \, d^3\bm x\right)^2}{2M_I}
\end{equation}
on the PPPS, all of whose states are again entangled between our charged balls and longitudinal photons.

And we again have, defining $\check H^{\mathrm{Schr}}_{\mathrm{QED}}$ by adding the same extra terms to $H^{\mathrm{Coulomb \, Schr}}_{\mathrm{QED}}$ that were added to $H^{\mathrm{Coulomb}}_{\mathrm{QED}}$ in \eqref{Hcheck},
\[
U\check H^{\mathrm{Schr}}_{\mathrm{QED}}U^{-1} = H^{\mathrm{PP\, Schr}}_{\mathrm{QED}},
\]
and
\[
U\bm p_I U^{-1} = \bm p_I - \int\hat{\bm A}^{\mathrm{long}}(\bm x)\rho_I(\bm x - \bm x_I)\, d^3x.
\]
etc.

In the usual approximation where we neglect terms which lead to radiative corrections (and suppress the
${\cal F}({\cal H}_{\mathrm{one}}^{\mathrm{trans}})\otimes$ in the Hilbert space) the Coulomb gauge formulation of this nonrelativistic version of QED gives us the familiar non-relativistic Schr\"odinger equation for systems of many charged particles.  (With extended balls, it is natural to include the [finite] self energies of the balls in $V_{\mathrm{Coulomb}}^{\mathrm{Schr}}$ and in $E$.)    So one might wonder how the product picture can work in the nonrelativistic limit since  the Hamiltonian of Equation \eqref{HamPPSchr} has no potential term!   To understand this, let us look at the familiar example of the (spinless) Hydrogen atom.  (For more details, see \cite[Section 4.2]{KayQES}.)  If Particle 1 is the proton and Particle 2 the electron and the two-body wave function $\Psi_{\mathrm{Schr}}$ is say a product of one of the usual Coulomb gauge energy eigenfunctions of the relative coordinate with a wave packet in the centre of mass coordinate of negligible energy, then in the usual Coulomb gauge description, we'll have
\[
\left(\frac{\bm p_1^2}{2M_1} + \frac{\bm p_2^2}{2M_2} + V_{\mathrm{Coulomb}}^{\mathrm{Schr}}\right)\Omega^{\mathrm{long}}\otimes \Psi_{\mathrm{Schr}} \approx E \Omega^{\mathrm{long}}\otimes \Psi_{\mathrm{Schr}}.
\]
In the product picture, this (on the CGPS) gets replaced by
\[
\left(\frac{\bm p_1^2}{2M_1} + \frac{\bm p_2^2}{2M_2} + {1\over 2}\mbox{$\tilde{\bm\pi}^{\mathrm{long}}$}^2\right )U\Omega^{\mathrm{long}}\otimes \Psi_{\mathrm{Schr}}  \approx E U\Omega^{\mathrm{long}}\otimes \Psi_{\mathrm{Schr}}.
\] 
(on the PPPS).  We see that the binding energy is now understood to be the \emph{decrease in the energy of the longitudinal part of the electric field when the charged balls are closer to one another}.  The (entangled) quantum state, $U\Omega^{\mathrm{long}}\otimes \Psi_{\mathrm{Schr}}$, schematically takes the form -- when we think of ${\cal F}({\cal H}_{\mathrm{one}}^{\mathrm{long}})\otimes L^2({\mathbb{R}}^6)$ as $L^2({\mathbb{R}}^6, \, {\cal F}({\cal H}_{\mathrm{one}}^{\mathrm{long}}))$ -- 
\[
({\bm x}_1, {\bm x}_2) \mapsto  \Psi_{\mathrm{Schr}}({\bm x}_1, {\bm x}_2)\Phi^{\mathrm{long}}({\bm x}_1, {\bm x}_2)
\]
where $\Phi^{\mathrm{long}}({\bm x}_1, {\bm x}_2)$ is the \emph{coherent state of longitudinal photons} which corresponds to the classical electric field due to the presence of the proton at ${\bm x}_1$ and the electron at ${\bm x}_2$.  

For more information about exactly what is meant by coherent states of longitudinal photons, see \cite{KayQES} which gives an alternative account of the product picture formulation of QED which takes the construction of such coherent states as its starting point.  We also refer the reader to the paper \cite{KayQES} for a discussion of a number of other topics including more about Gauss's law and the charged field commutator (see Section 3.3 there) and (in Section 4.3 there) about the reduced density operator of charged Schr\"odinger matter.   We shall also have more to say about the latter topic in the last section below.

\section{\label{Afterword} Afterword: The relevance to the matter gravity entanglement hypothesis and comments on the relation with Penrose's ideas about cosmological entropy and state vector reduction}

Aside from providing a new formulation of QED that is free from the aesthetically unpleasant features that we mentioned at the outset, and aside from offering the intriguing new way of thinking about the non-relativistic quantum mechanics of many charged particles that we discussed in the Section \ref{NonRelProd}, the product picture formulation of QED also permits us to ask some questions about QED that could not even be posed in the Coulomb gauge formulation.   In particular, thanks to the fact that, in the product picture, we can write the Hilbert space, ${\cal H}_{\mathrm{QED}}$, as a tensor product of the form \eqref{HilbProd}, it becomes meaningful to ask, for a (say pure) quantum state of QED -- thought of now as the density operator, $\sigma_{\mathrm{QED}}=|\bm\Psi\rangle\langle\bm\Psi|$ for $\bm\Psi \in {\cal H}_{\mathrm{QED}}$ -- what is its reduced density operator $\sigma_{\mathrm{charged \ matter}}$ on the Hilbert space ${\cal H}_{\mathrm{charged \ matter}}$ (which we define to be the partial trace of $\sigma$ over ${\cal H}_{\mathrm{electromag}}$).   Similarly, we could ask what is the reduced density operator $\sigma_{\mathrm{electromag}}$ on the Hilbert space ${\cal H}_{\mathrm{electromag}}$ (which we define to be the partial trace of $\sigma$ over ${\cal H}_{\mathrm{charged \ matter}}$).   And, relatedly, we can ask, e.g.\ what is the entanglement entropy of $\sigma_{\mathrm{QED}}$ between charged matter and the electromagnetic field.  Here we recall that, because of the assumed purity of $\sigma_{\mathrm{QED}}$, the von Neumann entropy of $\sigma_{\mathrm{charged \ matter}}$ and the von Neumann entropy of 
$\sigma_{\mathrm{electromag}}$ will be equal and their equal value is, deservedly named the charged matter-electromag entanglement entropy of $\sigma_{\mathrm{QED}}$. 

Aside from any possible interest in these questions in their own right, they are of interest as providing analogies to basic questions one asks on my \emph{matter gravity entanglement hypothesis} (see \cite{KayMatGrav,KayLeipzig} and references therein) which aims to help resolve a number of fundamental issues both about the foundations of quantum mechanics (in particular regarding the possibility of having an objective [observer-independent] notion of quantum state reduction) and also about the foundations of thermodynamics (in particular regarding the definition of entropy and the origin of the second law).   This hypothesis was inspired partly by the discovery \cite{Hawking} of the Hawking effect which suggests that there may be new interconnections between quantum mechanics, gravity and thermodynamics waiting to be discovered.   It was also party inspired by Roger Penrose's writings (see for example \cite{PenroseEinstVol}) on the status of the notion of entropy in a cosmological context and the importance of understanding how the entropy of the universe evolves in time from the big bang to the late-time universe.

The idea of the matter-gravity entanglement hypothesis is that, to address such questions, a closed system must be understood in the context of a (possibly low-energy approximate) theory of quantum gravity which is assumed, conservatively, to be formulated in terms of a total Hilbert space, say ${\cal H}_{\mathrm{total}}$  which arises as a tensor product,
${\cal H}_{\mathrm{gravity}}\otimes {\cal H}_{\mathrm{matter}}$ of a gravity Hilbert space ${\cal H}_{\mathrm{gravity}}$ and a matter Hilbert space ${\cal H}_{\mathrm{matter}}$.   It is further assumed that the state of a closed system is always a pure state, described by a density operator of form $\sigma_{\mathrm{total}}=|\bm\Psi\rangle\langle\bm\Psi|$ for $\bm\Psi \in H_{\mathrm{total}}$ and that (with a na\"ive notion of `time' and from a Schr\"odinger picture point of view) this evolves unitarily in time.   A fundamental process of decoherence (potentially offering an explanation of quantum state reduction) will then be expected to occur if we assume that the Hamiltonian generating our unitary time evolution (the quantum gravity Hamiltonian) arises  in the form (cf.\ Equation \eqref{HamPP})
\begin{equation}
\label{HamGrav}
H_{\mathrm{total}} = H_{\mathrm{gravity}} + H_{\mathrm{matter}} + H_{\mathrm{interaction}}
\end{equation}
and if we assume that the relevant density operator when we talk about the occurence of decoherence is not 
$\sigma_{\mathrm{total}}$ itself (which will always be pure) but rather the reduced density operator $\sigma_{\mathrm{matter}}$.   And we obtain a natural candidate for the physical entropy of a closed system by defining it to be the system's matter-gravity entanglement entropy (rather than the von Neumann entropy of $\sigma_{\mathrm{total}}$ which will always be zero).  In other words the von Neumann entropy of $\sigma_{\mathrm{matter}}$.   (Or equally of $\sigma_{\mathrm{gravity}}$.)

Indeed, suppose (now taking our closed system to be, say, the entire universe) we make the further assumption that the initial state was unentangled between matter and gravity (or perhaps just had a low degree of matter-gravity entanglement) then, for any of a wide range of interaction terms,  $H_{\mathrm{interaction}}$, in \eqref{HamGrav}, we would expect $\sigma_{\mathrm{matter}}$ to get more and more mixed (thus providing us with a fundamental process of decoherence) and concomitantly we would expect the matter-gravity entanglement entropy (i.e.\ our model for the physical entropy of the universe) to increase monotonically with time, thus offering the possibility of an objective explanation for the second law.

Moreover, taking instead our closed system to be a model asymptotically flat universe containing a ball of matter which collapses to a black hole, and similarly assuming that the initial  total state had a low degree of matter-gravity entanglement, we would again expect $\sigma_{\mathrm{matter}}$ to get more and more mixed and the system's matter-gravity entanglement entropy to increase monotonically.   Thus we would reconcile `information loss' (which we interpret now as the reduced density operator $\sigma_{\mathrm{matter}}$ getting more and more mixed) with an underlying unitary evolution of $\sigma_{\mathrm{total}}$.  

Thus this very simple hypothesis already appears to give us a conceptual template capable of both explaining  the second law and resolving the black hole information loss puzzle -- which, in fact, is seen now just to be a special case of the second law.

However, the entire hypothesis is predicated on quantum gravity having (to use the terminology of our introduction) a product picture.   And, even granting that quantum gravity (or at least an appropriate low-energy approximation of it which we might hope would be the suitable setting for discussing questions of decoherence and entropy) will turn out to be a conservative sort of (unitary) quantum theory, there might have seemed reasons to worry that this quantum theory would not have a product picture.   After all, one might have objected, quantum gravity is, in a certain well-known sense, a gauge theory and, more to the point, its Hamiltonian formulation involves four well-known constraints.   And one might then have argued by analogy that, just as the Gauss law constraint prevents quantum electrodynamics, in its Coulomb gauge formulation, from having a product picture, so one might expect that the four constraints of general relativity will prevent quantum gravity from having a product picture.   

However, now that we have provided QED with an alternative formulation which is a product picture, this worry is at least to some extent dispelled.   It raises the hope that, with further work, we will similarly be able to provide quantum gravity with a product picture.  

To end this afterword, I would like to make some tentative and sketchy remarks about how our matter gravity entanglement hypothesis might possibly be related to Roger Penrose's ideas (e.g.\ as discussed in \cite{PenroseEinstVol}) about cosmological entropy and especially to his related ideas (see e.g. \cite{Penrose1996}) about quantum state reduction.  It seems to me to be still quite unclear at present whether the two lines of thought are fully in contradiction with one another or whether our hypothesis perhaps stands in relation to Penrose's vision in somewhat the same way that, e.g., Bohr's 1910 theory of the Hydrogen atom related to the 1924 ideas of de Broglie (which led, a little later, to the Heisenberg-Schr\"odinger theory of quantum mechanics).

One point of contact is that both theories have addressed the issue of what happens to would-be macroscopic quantum superpositions involving different spacetime geometries and, for concreteness, both theories have been illustrated by considering a Schr\"odinger Cat-like superposition of two states of a single uniform density massive ball centred on two distinct locations in an otherwise empty (asymptotically flat) space.  Each ball state will have its own (Newtonian) gravitational field and the question is whether and in what sense the total state might spontaneously decohere.  According to Penrose's ideas \cite{Penrose1996} one expects such a macroscopic superposition to decay to a state centred on a single location on some time-scale which depends on the size and mass of the ball and the separation of the centres of mass of the two ball states in the superposition.   According to the matter gravity entanglement hypothesis, such a superposition could be in a static state but the relevant density operator, when one asks questions about purity versus mixedness, will not be its $\sigma_{\mathrm{total}}$ (which will always be pure) but rather it will be the reduced density operator $\sigma_{\mathrm{matter}}$.  One can compute this similarly to the way in which one would compute 
$\sigma_{\mathrm{charged \ matter}}$ in our product picture for QED for a superposition involving two states of an electrically charged ball centred on two different locations in the non-relativistic version of QED described in Section \ref{NonRelProd}.   In fact formulae for both $\sigma_{\mathrm{charged \ matter}}$ in the QED analogy and for $\sigma_{\mathrm{matter}}$ in weak-field Newtonian quantum gravity were obtained by the author in \cite{KayNewt} (albeit I was able, only around 24 years later, to provide them, in the case of QED, with the more satisfactory theoretical underpinning described here and in \cite{KayQES} [where the formulae are reproduced in Sections 1 and 4.3] and I should mention that work is in progress on providing such an underpinning [and the fixing of an overall constant] in the case of linearized gravity) and one finds that  (in the case the displacement of the centres is much smaller than the ball radius) the degree of mixedness of $\sigma_{\mathrm{charged\ matter}}$/$\sigma_{\mathrm{matter}}$ is governed by a certain \emph{decoherence length} which again depends on the size and mass of the ball and the separation of the centres of mass of the two ball states in the superposition.

As discussed in more detail in \cite{KayNewt} the formula obtained there for the decoherence length is remarkably similar to the formula obtained (up to a presently unfixable multiplicative constant) by Penrose in \cite{Penrose1996} for his decay time-scale.   (The only difference is that where there is a Laplacian in the latter formula, there is a square root of the Laplacian in the former.)   On the other hand there are important physical differences.  One difference, as we have already seen, is that the decoherence is a time-dependent process in the Penrose theory while it is ever present in the setting of \cite{KayNewt}. Another is that the sort of decoherence one obtains on the theory of \cite{KayQES,KayNewt} is, in the case of static or slowly changing ball configurations, a reversible sort of decoherence; if one were to (slowly) bring the ball centres back into coincidence, the decohered ball states would recohere.   For this reason, in the Newtonian limit, the matter gravity entanglement hypothesis appears (see \cite{KayAbyaneh} for details) to predict the same result as standard quantum mechanics for the experiment \cite{Bouw} proposed by Penrose and thus to be experimentally distinguishable from his theory.   One expects instead that, on the matter-gravity entanglement hypothesis, irreversible decoherence would occur and lead to a noticeable loss of coherence in that experiment only if the oscillator in the experiment were to oscillate sufficiently rapidly to emit a graviton in a suitably short time interval -- a much harder regime to access.   (For more discussion and for more discussion of the physical interpretation of $\sigma_{\mathrm{matter}}$ see again \cite{KayAbyaneh} and \cite{KayLeipzig}.)

Another point of difference between the two sets of ideas is that, on the matter-gravity entanglement hypothesis, the entropy of the universe resides at all times in the entanglement of matter with gravity (which is assumed to start low and then predicted to increase) whereas, on Penrose's ideas, one apparently thinks in terms of an entropy which is a sum of a matter entropy (which dominates at early times) and a gravity entropy (which dominates at late times and is larger than the matter entropy at early times).   Here there appears to be an antagonism between the two theories which resembles (and partly subsumes) the antagonism between the understanding of black hole entropy on the matter-gravity entanglement hypothesis (according to which the total state is pure while it is the reduced density operator of matter as well as the reduced density operator of gravity which are approximately Gibbs states) and the traditional (Hawking) understanding of black holes according to which the total state is a (mixed) Gibbs state.  See the proposed resolution of the \emph{thermal atmosphere puzzle} in \cite{KayMatGrav}. 

\section*{Acknowledgments}

The author thanks the Leverhulme Foundation for the award of Leverhulme Fellowship 
\hfil\break
RF\&G/9/RFG/2002/0377 for the period October 2002 to June 2003 during which some of this work was done. I thank Jonathan Dimock  and Michael Kay for valuable comments.

\section*{Data Availability}

Data sharing is not applicable to this article as no new data were created or analyzed in this
study.

\section*{Author Declaration}

The author has no conflicts to disclose.

\end{document}